\documentclass[journal]{IEEEtran}

\usepackage[dvipsnames]{xcolor}
\usepackage[acronym]{glossaries}
\usepackage[colorlinks, citecolor=blue, linkcolor=blue]{hyperref}
\usepackage{todonotes}
\usepackage{comment}
\usepackage{nameref}
\usepackage{subcaption}
\usepackage{multicol}
\usepackage{multirow}
\usepackage{rotating}
\usepackage{stfloats}
\usepackage[noadjust]{cite}
\usepackage{float}
\usepackage{amsfonts}
\usepackage{upgreek}
\usepackage{mathtools}
\usepackage{soul}
\usepackage{tcolorbox}
\newcommand{\SNR}{\mathrm{SNR}}
\newcommand{\dB}{\mathrm{dB}}
\newcommand{\mdg}{\mathrm{mdg}}
\newcommand{\lnL}[1]{\mathrm{ln}\left(#1\right)}
\newcommand{\logL}[2]{\mathrm{log}_{#1}\left(#2\right)}

\newcommand{\sinL}[1]{\mathrm{sin}\left(#1\right)}
\newcommand{\sinPowL}[2]{\mathrm{sin}^{#1}\left(#2\right)}
\newcommand{\cosL}[1]{\mathrm{cos}\left(#1\right)}
\newcommand{\cosPowL}[2]{\mathrm{cos}^{#1}\left(#2\right)}
\newcommand{\arcsinL}[1]{\mathrm{sin^{-1}}\left(#1\right)}
\newcommand{\cov}[1]{\mathrm{cov}\left(#1\right)}

\newcommand{\TabularCoefficients}{
    \begin{tabular}{cc|cc|cc|cc|cc|}
        \cline{3-10}
         &  & \multicolumn{2}{c|}{$\SNR=5\ \dB$} & \multicolumn{2}{c|}{$\SNR=10\ \dB$} & \multicolumn{2}{c|}{$\SNR=15\ \dB$} & \multicolumn{2}{c|}{$\SNR=20\ \dB$} \\ \cline{2-10}
         \multicolumn{1}{l|}{} & $D$ & $\gamma_{0}$ & $\gamma_{1}$ & $\gamma_{0}$ & $\gamma_{1}$ & $\gamma_{0}$ & $\gamma_{1}$ & $\gamma_{0}$ & $\gamma_{1}$ \\ \hline
    \multicolumn{1}{|c|}{\multirow{7}{*}{\begin{sideways}Method \ref{ssec:gue_spectral_distribution}\end{sideways}}}
                           & 2 & $2.16297989$ & $-1.424485\mathrm{E-}3$ & $2.16381222$ & $-3.241463\mathrm{E-}4$ & $2.16393724$ & $-5.664255\mathrm{E-}5$ & $2.16395171$ & $-1.031541\mathrm{E-}5$ \\
    \multicolumn{1}{|c|}{} & 3 & $1.09160893$ & $~~6.095528\mathrm{E-}4$ & $1.09153003$ & $~~1.498233\mathrm{E-}4$ & $1.09149336$ & $~~4.022044\mathrm{E-}5$ & $1.09147997$ & $~~7.691563\mathrm{E-}6$ \\
    \multicolumn{1}{|c|}{} & 4 & $0.72791071$ & $~~1.114647\mathrm{E-}4$ & $0.72795167$ & $~~5.751291\mathrm{E-}5$ & $0.72793629$ & $~~2.789583\mathrm{E-}5$ & $0.72792665$ & $~~6.419168\mathrm{E-}6$ \\
    \multicolumn{1}{|c|}{} & 5 & $0.54480425$ & $-1.547994\mathrm{E-}4$ & $0.54492813$ & $-2.174217\mathrm{E-}5$ & $0.54493125$ & $~~1.038573\mathrm{E-}5$ & $0.54492585$ & $~~5.543675\mathrm{E-}6$ \\
    \multicolumn{1}{|c|}{} & 6 & $0.43511438$ & $~~7.630925\mathrm{E-}5$ & $0.43513127$ & $~~3.758373\mathrm{E-}5$ & $0.43512306$ & $~~1.614824\mathrm{E-}5$ & $0.43511820$ & $~~3.966745\mathrm{E-}6$ \\
    \multicolumn{1}{|c|}{} & 7 & $0.36203898$ & $-1.120187\mathrm{E-}4$ & $0.36211666$ & $-1.754819\mathrm{E-}5$ & $0.36212190$ & $~~4.308864\mathrm{E-}6$ & $0.36212014$ & $~~3.102686\mathrm{E-}6$ \\
    \multicolumn{1}{|c|}{} & 8 & $0.31008496$ & $-1.768119\mathrm{E-}4$ & $0.31016253$ & $-3.614410\mathrm{E-}5$ & $0.31017045$ & $-2.168861\mathrm{E-}6$ & $0.31017007$ & $~~1.195690\mathrm{E-}6$ \\ \hline
    \multicolumn{1}{|c|}{\multirow{6}{*}{\begin{sideways}Method \ref{ssec:wigner_semicircular_distribution}\end{sideways}}}
                     & 9 & $0.27114473$ & $~~1.871205\mathrm{E-}4$ & $0.27107389$ & $~~3.189872\mathrm{E-}5$ & $0.27106495$ & $~~1.572890\mathrm{E-}6$ & $0.27106446$ & $-4.804128\mathrm{E-}7$ \\
    \multicolumn{1}{|c|}{} & 10 & $0.24115580$ & $~~8.250165\mathrm{E-}5$ & $0.24113560$ & $~~3.632280\mathrm{E-}6$ & $0.24113577$ & $-4.177426\mathrm{E-}6$ & $0.24113686$ & $-1.934882\mathrm{E-}6$ \\
    \multicolumn{1}{|c|}{} & 12 & $0.19777600$ & $~~7.311636\mathrm{E-}5$ & $0.19775961$ & $-1.298120\mathrm{E-}7$ & $0.19776223$ & $-7.621126\mathrm{E-}6$ & $0.19776444$ & $-3.509853\mathrm{E-}6$ \\
    \multicolumn{1}{|c|}{} & 20 & $0.11577972$ & $-2.081275\mathrm{E-}4$ & $0.11585818$ & $-7.543138\mathrm{E-}5$ & $0.11587780$ & $-2.177745\mathrm{E-}5$ & $0.11588300$ & $-6.367806\mathrm{E-}6$ \\
    \multicolumn{1}{|c|}{} & 30 & $0.07681464$ & $-2.282790\mathrm{E-}4$ & $0.07689957$ & $-8.327005\mathrm{E-}5$ & $0.07691975$ & $-2.324244\mathrm{E-}5$ & $0.07692502$ & $-6.536148\mathrm{E-}6$ \\
    \multicolumn{1}{|c|}{} & 40 & $0.05768991$ & $-1.842371\mathrm{E-}4$ & $0.05776636$ & $-6.941402\mathrm{E-}5$ & $0.05778465$ & $-2.036533\mathrm{E-}5$ & $0.05778933$ & $-5.994924\mathrm{E-}6$ \\ \hline
    \end{tabular}
}
\newcommand{\appref}[1]{\hyperref[#1]{appendix~\ref*{#1}}}
\newacronym{snr}{SNR}{signal-to-noise ratio}
\newacronym{sdm}{SDM}{space-division multiplexing}
\newacronym{wdm}{WDM}{wavelength-division multiplexing}
\newacronym{pdg}{PDG}{polarization-dependent gain}
\newacronym{pdl}{PDL}{polarization-dependent loss}
\newacronym{mdg}{MDG}{mode-dependent gain}
\newacronym{mdl}{MDL}{mode-dependent loss}
\newacronym{csi}{CSI}{channel state information}
\newacronym{cdf}{CDF}{cumulative distribution function}
\newacronym{pdf}{PDF}{probability density function}
\newacronym{gue}{GUE}{Gaussian unitary ensemble}
\newacronym{mmse}{MMSE}{minimum mean squared error}
\newacronym{msle}{MSLE}{mean squared logarithmic error}
\newacronym{mux}{MUX}{multiplexer}
\newacronym{demux}{DEMUX}{demultiplexer}
\newacronym{mimo}{MIMO}{multiple-input multiple-output}
\newacronym{awgn}{AWGN}{additive white Gaussian noise}
\newacronym{dsp}{DSP}{digital signal processing}

\begin{document}
\title{Analytic Models for the Capacity Distribution in MDG-impaired Optical SDM Transmission}
\author{Lucas~Alves~Zischler,~\IEEEmembership{Student~Member,~IEEE,}
        and~Darli~A.~A.~Mello,~\IEEEmembership{Member,~IEEE}%
% Lower left notes
\thanks{Manuscript received August 24, 2024; revised 21 December 2024 and 25 February 2025; accepted 14 March 2025. This work was financed in part by the Coordenação de Aperfeiçoamento de Pessoal de N\'{i}vel Superior – Brasil (CAPES) – Finance Code 001, and by Fapesp grant 2022/11596-0.}%
\thanks{L. Alves Zischler, and D. A. A. Mello are with the School of Electrical and Computer Engineering, State University of Campinas, Campinas 13083-970, Brazil: (e-mail: l257176@dac.unicamp.br).}}%

% Paper headers
\markboth{}%
{Analytic Models for the Capacity Distribution in MDG-impaired Optical SDM Transmission}

% Title area
\maketitle

% Abstract
\begin{abstract}
In coupled space-division multiplexing (SDM) transmission systems, imperfections in optical amplifiers and passive devices introduce mode-dependent loss (MDL) and gain (MDG). These effects render the channel capacity stochastic and result in a decrease in average capacity. Several previous studies employ multi-section simulations to model the capacity of these systems. Additionally, relevant works derive analytically the capacity distribution for a single-mode system with polarization-dependent gain and loss (mode count $D=2$). However, to the best of our knowledge, analytic expressions of the capacity distribution for systems with $D>2$ have not been presented. In this paper, we provide analytic expressions for the capacity of SDM optical systems affected by MDG with arbitrary mode counts. The expressions rely on Gaussian approximations for the per-mode capacity distributions and for the overall capacity distribution, as well as on fitting parameters for the capacity cross-correlation among different modes. Compared to simulations, the derived analytical expressions exhibit a suitable level of accuracy across a wide range of practical scenarios.
\end{abstract}

% Keywords

% Glossary
\glsreset{snr}
\glsreset{sdm}
\glsreset{mdg}
\glsreset{cdf}
\glsreset{pdf}
\glsreset{gue}
\glsreset{mmse}
\glsreset{msle}

% Section: Introduction
\section{Introduction}

\IEEEPARstart{T}{he increased} capacity of multi-mode or multi-core fibers is a field of active research to handle the growing data traffic~\cite{morioka2009new, essiambre2012capacity,winzer2012optical,richardson2013space,saitoh2016multicore,puttnam2021space, cristiani2022roadmap,rademacher202010, rademacher2021peta, rademacher20221}. However, unequal gains in amplifiers or unequal losses in passive devices and fibers generate an effect known as \gls*{mdg} and \gls*{mdl}. In deployed fibers, the \gls*{mdg} or \gls*{mdl} generates a stochastic effect that turns the channel capacity into a random variable~\cite{winzer2014mode}. The stochastic capacity in \gls*{sdm} channels introduces a probability of outage if the receiver is set to operate on data rates higher than those instantaneously supported by the channel~\cite{winzer2011mimo}. The effects of a stochastic capacity have been extensively addressed in the past in the scope  of \gls*{pdl}~\cite{yamamoto1993observation,taylor1993observation,pilipetskii2006performance,shtaif2004polarization}. While the loss in current optical devices for single-mode systems has small polarization dependency~\cite{nelson2011statistics}, current \gls*{sdm} devices have significant levels of \gls*{mdg} and \gls*{mdl} that cannot be neglected~\cite{velazquez2018scaling,wada2018recent,wada2018cladding}. Communications systems with a stochastic \gls*{snr} need to be designed to maintain the outage probability below accepted levels. To properly assess outages in the design of future \gls*{sdm} systems, statistical models for the capacity are necessary.

The capacity loss statistics for a set of amplification schemes are discussed in~\cite{antonelli2015modeling}. Expressions for the variance of the per-mode capacity loss for four different amplification schemes are derived as a function of the channel \gls*{mdl} vector, discussed in~\cite{antonelli2019stokes}. A closed-form expression for the average capacity is presented in \cite{ho2011mode}, but the capacity variance is not calculated. The impact of frequency diversity on the capacity distribution has been discussed in~\cite{ho2011frequency}, but no analytical expression for the capacity distribution has been presented. The exact distribution for the capacity of a single-mode-fiber optical system with \gls*{pdg} ($D=2$) has been presented in~\cite{mello2020impact}. However, the statistics of \gls*{sdm} systems with higher mode counts have been addressed by simulation.

In this paper, we provide analytical expressions for the variance of the capacity distribution\footnote{An analytical solution for the total capacity mean is presented in~\cite[Eq.~(10)]{ho2011mode}.}. The derivations are based on the assumption of a Gaussian distribution for the per-mode and total capacity\footnote{The Gaussian assumption for the total capacity distribution has been discussed in~\cite{melo2024on}, confirming its validity for increasing mode counts.} \glspl*{pdf}. The joint distribution of per-mode capacities is approximated as a multivariate normal, and the total capacity is obtained as the linear combination of the per-mode capacities. The derivations rely on the modal gain distributions outlined in \cite{ho2011mode}.

The remainder of this work is structured as follows. Section~\ref{sec:instcap} reviews the channel capacity in strongly-coupled \gls*{sdm} systems subject to \gls*{mdg} and explains the most relevant statistical distributions that are investigated throughout the paper. Section~\ref{sec:marginaldistribution} derives the statistical distribution of the per-mode capacity using two different methods. Section~\ref{sec:correlation} uses the previously defined per-mode capacity distribution to obtain the total capacity distribution. Section~\ref{sec:freqdiv} investigates the effects of frequency diversity on the capacity variance and validates the derived analytical model. Section~\ref{sec:casestudy} presents a step-by-step application of the derived model to estimate the outage capacity of a sample case study. Lastly, Section \ref{sec:conclusion} concludes the paper.

% Section: 1
\section{Capacity Metrics In Optical SDM}
\label{sec:instcap}

\begin{figure*}[!ht]
    \centering
    \includegraphics{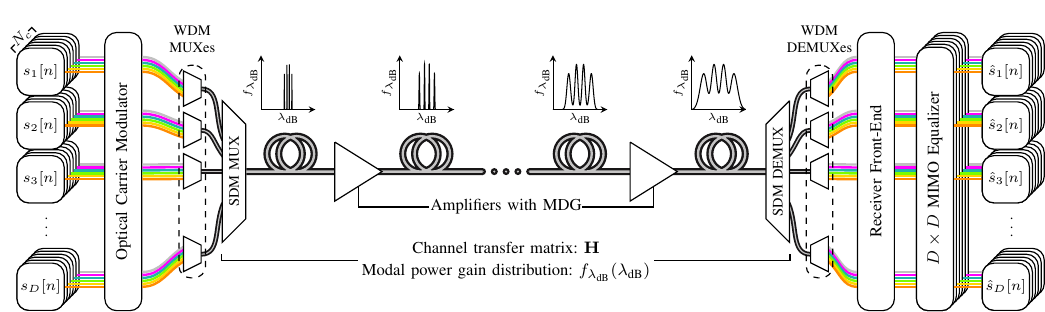}
    \caption{Representation of a coupled \gls*{sdm} system with $N_{c}$ spectral channels and $D$ spatial channels. The spectral and spatial channels are first modulated. Subsequently, a spectral multiplexer generates a WDM signal, and a spatial multiplexer generates an SDM signal.  The transmitted symbol sequences $s_{i}[n]$, for $i\in[1,D]$, are recovered independently for each spectral channel by a $D\times D$ MIMO equalization process, generating estimates $\widehat{s}_{i}[n]$ of the transmitted symbol sequence. The optical carrier modulator and the receiver front-end perform electrical-to-optical and optical-to-electrical conversion, respectively. The channel transfer matrix $\mathbf{H}$ and modal gains $\lambda_{\dB}$ are frequency dependent, but we omit the frequency dependent variable for simplicity, assuming a narrowband channel. The insets illustrate the evolution of the logarithmic scale modal gains \gls*{pdf} ($f_{\lambda_{\text{dB}}}(\lambda_{\text{dB}})$).}
    \label{fig:Diagram}
\end{figure*}

We consider the \gls*{wdm} and \gls*{sdm} system  presented in Fig.~\ref{fig:Diagram}. The $N_c$-channel spectral multiplexer generates the WDM signal and the $D$-channel spectral multiplexer generates the SDM signal. In the mathematical derivations, we analyze only a narrowband channel, i.e., a channel with flat frequency response along the signal bandwidth. $D$ independent signals $s_i$, $i \in[1,D]$, are spatially multiplexed into the \gls*{sdm} fiber system for each frequency channel. The \gls*{sdm} signal is amplified multiple times by amplifiers with \gls*{mdg} that maintain a total output power evaluated across all spatial and spectral channels. Assuming a linear channel, transmission can be represented by a transfer matrix $\mathbf{H}$ followed by additive noise at the receiver \cite{ho2011mode}. The coupled signals are then separated at the receiver by a $D\times D$ \gls*{mimo} \gls*{dsp} structure.

Assuming that no \gls*{csi} is available at the transmitter, the total instantaneous capacity of a narrowband \gls*{sdm} channel is given by~\cite{paulraj2003introduction}
\begin{equation}
    C_{T}=\logL{2}{\mathrm{det}\left[\mathbf{I}+\SNR\cdot\mathbf{H}\mathbf{H}^{H}\right]},
\end{equation}
where \gls*{snr} is the ratio of total signal power and total noise power, $\mathbf{H}$ is the channel transfer matrix, and $\mathbf{I}$ is the identity matrix of size $D$, where $D$ is the mode count. We assume a narrowband channel for most of the work. A discussion on the frequency dependence of the modal gains is provided in Section~\ref{sec:freqdiv}.
The total instantaneous capacity can also be obtained from the linear scale eigenvalues $\lambda$  of the $\mathbf{H}\mathbf{H}^{H}$~\cite{paulraj2003introduction} operator
\begin{equation}
	C_{T} = \sum_{i=1}^{D} C_{i} = \sum_{i=1}^{D} \logL{2}{1+\SNR\cdot\lambda_{i}},
	\label{eq:capacity_total}
\end{equation}
where $\lambda_{i}$ is the $i^{th}$ Schmidt mode power gain\footnote{The Schmidt modes are the optimal basis of propagation. For a discussion on Schmidt modes, see~\cite{juarez2012perspectives, ho2013mode}.}. We consider noise as spatially white\footnote{The spatial whiteness assumption has stronger validity for higher mode counts and wider bandwidths~\cite{efimov2014spatial}. The expressions presented in~\cite{antonelli2015modeling} also account for other amplification regimes.}, yielding equal noise power for all modes \cite{ho2011mode}. Assuming that \gls*{sdm} amplifiers maintain a constant total power after amplification, the sum of all $\lambda_i$ is set to $D$.

As shown in (\ref{eq:capacity_total}), the total capacity $C_T$ delivered by the system is the sum of the capacities of the $D$ individual Schmidt modes. The eigenvalues $\lambda_i$ in (\ref{eq:capacity_total}) can be interpreted as a relative multiplication factor that scales the \gls*{snr} of the system for each mode. Without \gls*{mdg}, all $\lambda_i$s are the same. With \gls*{mdg}, they are different for each mode and vary over time. Let us now return to our primary question, which is to determine the distribution $f_{C_T}$ of the total capacity $C_T$. Having access to $f_{C_T}$ enables the system designer to avoid outages that could disrupt the system while minimizing resource overprovisioning. 
Clearly, $C_T$ depends on the distribution $f_{C_i}$ of the individual modes. The following derivations demonstrate that the distribution $f_{C_i}$ of each mode can be approximated as Gaussian. Consequently, as $C_T$ is the sum of the individual capacities $C_i$, it can also be assumed to follow a Gaussian distribution. Assuming both $C_T$ and $C_i$ as Gaussian, the distribution can be completely described by the mean and standard deviation. While the mean of $C_T$ has been previously derived in \cite{ho2011frequency}, its variance is derived for the first time in this paper. We first derive approximations for distribution $f_{\lambda_{\textrm{dB},i}}$ of $\lambda_i$, expressed in dB. Then, from $f_{\lambda_{\textrm{dB},i}}$ we obtain $f_{C_i}$ using properties of distribution transformation. Finally, assuming $f_{C_i}$ as Gaussian, we obtain a Gaussian $f_{C_T}$. This step uses conventional properties of sums of correlated Gaussian random variables.

\begin{figure*}[!ht]
    \centering
    \includegraphics{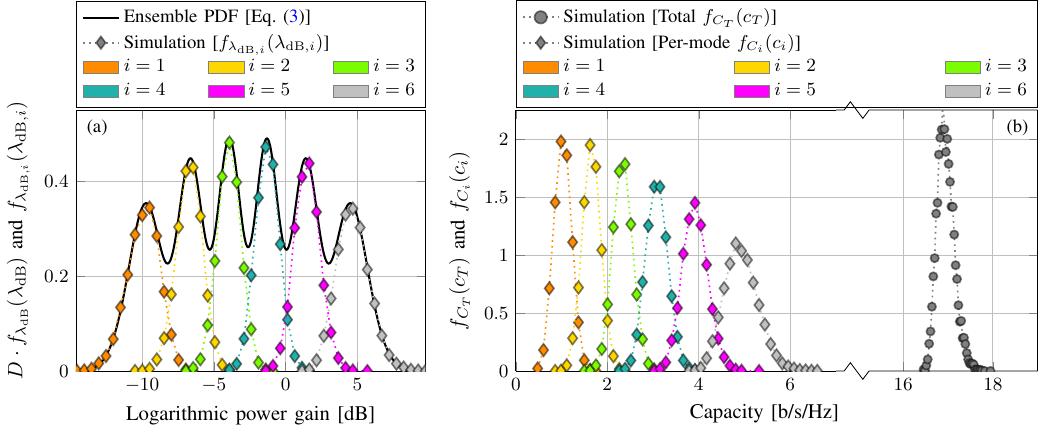}
    \caption{(a) Logarithmic gain $\lambda_{\dB}$ distribution of each mode and ensemble, considering the sum of the linear gains equals to $D$. The ensemble distribution is normalized by $D$ in order to facilitate visual analysis. The per-mode distributions are obtained via simulation for ${D=6}$ and ${\sigma_{\mdg}=5~\dB}$. (b) Instantaneous capacity distribution of each mode and total. The distributions are obtained via simulation for ${D=6}$, ${\SNR=10~\dB}$, and ${\sigma_{\mdg}=5~\dB}$. Dotted lines connect the simulated histogram values for visualization purposes.}
    \label{fig:InitialDistributions}
\end{figure*}

Fig.~\ref{fig:InitialDistributions} illustrates the most important distributions addressed in this paper. Fig.~\ref{fig:InitialDistributions}a presents the distribution of the power gains $\lambda_i$ in logarithmic scale. The distributions for the individual $\lambda_i$s, shown by markers, are obtained by Monte-Carlo simulations\footnote{The simulation results rely on the multisection model discussed in~\cite{ho2011mode}. The per-section \gls*{mdg} is controlled such that a target $\sigma_{\mdg}$ is achieved. From the channel matrix, we obtain the instantaneous modal gains and capacity values. The statistical distribution and values are obtained over 100 Monte-Carlo realizations. The results are obtained by simulating 100 sections with 50 km each. The mode count, \gls*{snr}, and~$\sigma_{\mdg}$ values of each simulation are specified on the respective figures.}. The solid black curve represents the analytical distribution of the ensemble of power gains $f_{\lambda_{\textrm{dB}}}$ as presented in \cite{ho2011mode}, normalized by the number of modes, and is included here for reference. In a later section, we derive an analytical approximation for this distribution for the first time. Fig.~\ref{fig:InitialDistributions}b illustrates the distribution of the per-mode capacity $C_i$ and the total capacity $C_T$. Again, these distributions are obtained through simulations, and analytical expressions are derived in later sections. It is interesting to see how the different modes contribute to different capacities. The figures also suggest that the distribution of $C_i$ and $C_T$ can be approximated as Gaussian.

% Section: 2
\section{Per-Mode Capacity Distribution $C_i$}
\label{sec:marginaldistribution}

We evaluate the statistics of the total capacity $C_{T}$ based on the per-mode individual $\lambda_{i}$ and $C_{i}$ distributions. Approximations on $C_{i}$ distributions are obtained from two methods derived from the \gls*{gue} spectral and Wigner semicircular distributions for $\lambda_{i}$. 

\subsection{Approximations on the $C_{i}$ distribution based on the GUE spectral distribution for $\lambda_{i}$}
\label{ssec:gue_spectral_distribution}

The \gls*{pdf} $f_{\lambda_{\dB}}(\lambda_{\dB})$ of the $\lambda_{\dB}$ eigenvalues for a $D^{th}$ order \gls*{gue} is given by~\cite[Table 1]{ho2011mode}, \cite[Eq.~(S.15)]{ho2011statistics}
\begin{equation}
	\begin{split}
		f_{\lambda_{\dB}}(\lambda_{\dB}) = &\frac{\alpha_{\lambda_{\dB},D}}{\sigma_{\mdg}}e^{-\frac{(D+1)}{2}\left(\frac{\lambda_{\dB}-\mu_{\lambda_{\dB}}}{\sigma_{\mdg}}\right)^{2}}\\
		&\sum_{k=0}^{D-1}\beta_{\lambda_{\dB},D,k}\left(\frac{\lambda_{\dB}-\mu_{\lambda_{\dB}}}{\sigma_{\mdg}}\right)^{2k}
	\end{split}
	\label{eq:flambdadb}
\end{equation}
where $\alpha_{\lambda_{\dB},D}$ is a normalization factor such that the \gls*{pdf} integral equals 1, and $\beta_{\lambda_{\dB},D,k}$ are polynomial coefficients. $\alpha_{\lambda_{\dB},D}$ and $\beta_{\lambda_{\dB},D,k}$ are constant for a given number of modes. For $D\le8$ the values of $\alpha_{\lambda_{\dB},D}$ and $\beta_{\lambda_{\dB},D,k}$ are provided in \cite[Table~1]{ho2011mode}, and the derivation of the coefficients is presented in \appref{app:a}. Values\footnote{The notation $\sigma_{\mdg}$ was preferred to $\sigma_{\lambda_{\dB}}$, as the former is predominant in literature.} $\sigma_{\mdg}$ and $\mu_{\lambda_{\dB}}$ are the standard deviation and the mean of $f_{\lambda_{\dB}}(\lambda_{\dB})$, respectively. Value  $\mu_{\lambda_{\dB}}$ is given such that the linear scale eigenvalue distribution $f_{\lambda}(\lambda)$ has unitary mean, and is given by (see \appref{app:b})
\begin{equation}
    \mu_{\lambda_{\dB}}=-10\cdot\mathrm{log}_{10}\left[\int_{-\infty}^{\infty}10^{\frac{x}{10}}\widehat{f}_{\lambda_{\dB}}(x)dx\right],
    \label{eq:mulambdadb}
\end{equation}
 where $\widehat{f}_{\lambda_{\dB}}(x)$ is the zero mean $\lambda_{\dB}$ distribution.

We assume that the means of individual eigenvalues $\lambda_{\dB,i}$  are near the local maxima of $f_{\lambda_{\dB}}$. We further assume that  $\lambda_{\dB,i}$ is Gaussian-distributed with standard deviation $\sigma_{\lambda_{dB,i}}$ and mean $\mu_{\lambda_{\dB,i}}$. The ordering of the indices is given by their mean value such that $\mu_{\lambda_{\dB,1}}<\mu_{\lambda_{\dB,2}}<\cdots<\mu_{\lambda_{\dB,D}}$.

We approximate $\mu_{\lambda_{\dB,i}}$  by the odd-index zeros of the $f_{\lambda_{\dB}}$ derivative (for a detailed derivation, see \appref{app:c}), yielding
\begin{equation}
    \begin{split}
        \sum_{k=0}^{D-1}\frac{\beta_{\lambda_{\dB},D,k}}{\sigma_{\mdg}^{2k}}\left[2k\cdot\left(\mu_{\lambda_{\dB,i}}-\mu_{\lambda_{\dB}}\right)^{2k-1}\right. \\
        \left. -\frac{D+1}{\sigma_{\mdg}^{2}}\left(\mu_{\lambda_{\dB,i}}-\mu_{\lambda_{\dB}}\right)^{2k+1}\right] &= 0. 
    \end{split}
    \label{eq:ulambdaguederivative}
\end{equation}

\begin{figure*}[!ht]
    \centering
    \includegraphics{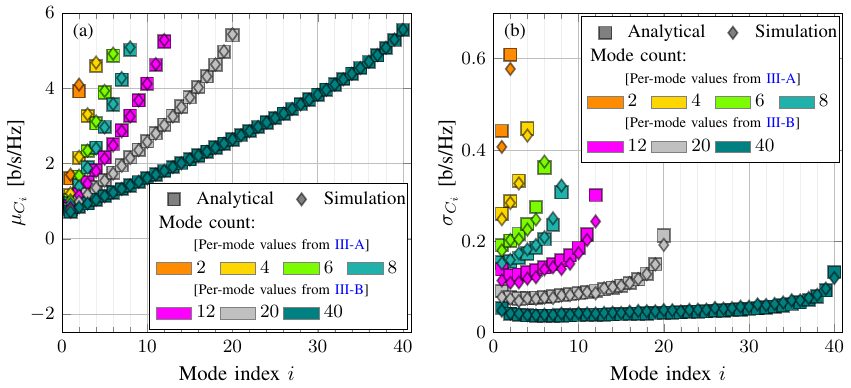}
    \caption{Analytical approximation of per-mode capacity mean (a) and standard deviation (b) for ${\SNR=10~\dB}$ and ${\sigma_{\mdg}=5~\dB}$. For ${D\leq 8}$ the \gls*{gue} distribution method is considered. Higher mode counts use the Wigner semicircular \gls*{cdf} method.}
    \label{fig:Marginals}
\end{figure*}

\begin{figure}[!h]
    \centering
    \includegraphics{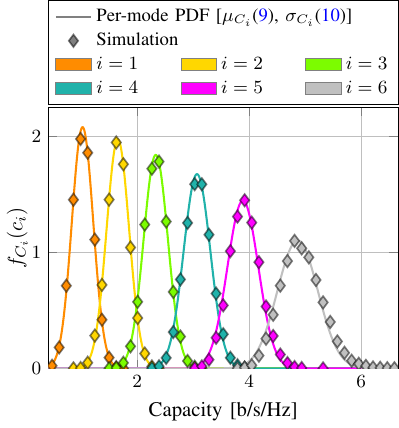}
    \caption{Comparison between the analytically obtained curves, assuming a Gaussian distribution, and simulation values for the per-mode capacity \gls*{pdf}s for ${D=6}$, ${\SNR=10~\dB}$, and ${\sigma_{\mdg}=5~\dB}$.}
    \label{fig:MargHist}
\end{figure}

The values of $\sigma_{\lambda_{\dB,i}}$ are obtained by making the ensemble and individual \gls*{pdf} values equal at $\mu_{\lambda_{\dB,i}}$ and applying a normalization factor of $1/D$
\begin{equation}
	\sigma_{\lambda_{\dB,i}} = \frac{1}{D\sqrt{2\pi} f_{\lambda_{\dB}}(\mu_{\lambda_{\dB,i}})}.
    \label{eq:sigmalambdadb}
\end{equation}

From the obtained $\mu_{\lambda_{\dB,i}}$ and $\sigma_{\lambda_{\dB,i}}$, the per-mode \gls*{pdf} is approximated by a Gaussian distribution
\begin{equation}
    f_{\lambda_{\dB,i}}(\lambda_{\dB,i}) = \frac{1}{\sigma_{\lambda_{\dB,i}}\sqrt{2\pi}}e^{-\frac{1}{2}\left(\frac{\lambda_{\dB,i}-\mu_{\lambda_{\dB,i}}}{\sigma_{\lambda_{\dB,i}}}\right)^{2}}.
    \label{eq:flambdadbi}
\end{equation}

The relation between the individual eigenvalue distribution $f_{\lambda_{\dB,i}}$ and  its corresponding capacity distribution $f_{C_{i}}(c_{i})$ is obtained by transformation as
\begin{equation}
    f_{C_{i}}(c_{i}) = \frac{10\cdot\lnL{2}2^{c_{i}}}{\lnL{10}\left(2^{c_{i}}-1\right)}f_{\lambda_{\dB,i}}\left(\frac{10}{\lnL{10}}\lnL{\frac{2^{c_{i}}-1}{\SNR}}\right).
    \label{eq:fclambdadbi}
\end{equation}

We also approximate the $C_{i}$ \gls*{pdf} by a Gaussian distribution with standard deviation $\sigma_{C_{i}}$ and mean $\mu_{C_{i}}$. Setting the derivative of \eqref{eq:fclambdadbi} to zero (see \appref{app:d}), the value of $\mu_{C_{i}}$ can be obtained as the solution of the simplified equation
\begin{equation}
	\frac{10\cdot2^{\mu_{C_{i}}}}{\sigma_{\lambda_{\dB,i}}^{2}\lnL{10}}\left(\frac{10}{\lnL{10}}\lnL{\frac{2^{\mu_{C_{i}}}-1}{\SNR}}-\mu_{\lambda_{\dB,i}}\right)+1 = 0.
    \label{eq:mucigue}
\end{equation}

The standard deviation values $\sigma_{C,i}$ are obtained by assuming a Gaussian approximation and setting \eqref{eq:fclambdadbi} equal at $\mu_{C,i}$, which is achieved for
\begin{equation}
	\sigma_{C_{i}} = \frac{1}{\sqrt{2\pi} f_{C_{i}}(\mu_{C_{i}})}.
	\label{eq:sigmacigue}
\end{equation}

An evaluation of $\mu_{C_i}$ and $\sigma_{C_i}$ for $2 \leq D \leq 8$ is presented in Fig.~\ref{fig:Marginals}. The accuracy of the Gaussian approximation for the individual capacity distributions $f_{C_{i}}$, for a system with $D=6$, $\SNR=10~\dB$ and $\sigma_{\mdg} = 5~\dB$, is shown in Fig.~\ref{fig:MargHist}. The results reveal, again, a suitable agreement between simulation and the derived analytical curves.

\subsection{Approximations on the $C_{i}$ distribution based on the Wigner semicircular distribution for $\lambda_{i}$}
\label{ssec:wigner_semicircular_distribution}

The limiting distribution of $\lambda_{\dB}$ as $D\rightarrow\infty$ is given by the Wigner semicircular distribution as~\cite[I]{ wigner1955characteristic},~\cite[Table 1]{ho2011mode}
\begin{equation}
    \begin{gathered}
	    f_{\lambda_{\dB}}(\lambda_{\dB}) = \frac{1}{2\pi\sigma_{\mdg}}\sqrt{4-\frac{\left(\lambda_{\dB}-\mu_{\lambda_{\dB}}\right)^{2}}{\sigma_{\mdg}^{2}}},\\
        -2\sigma_{\mdg}+\mu_{\lambda_{\dB}}\leq\lambda_{\dB}\leq 2\sigma_{\mdg}+\mu_{\lambda_{\dB}},
    \end{gathered}
    \label{eq:flambdadbwigner}
\end{equation}
with $\mu_{\lambda_{\dB}}$ given in \eqref{eq:mulambdadb}.

For this limiting case, the capacity \gls*{pdf} and \gls*{cdf} for the ensemble of eigenvalues can be derived analytically. The capacity \gls*{pdf} is obtained applying \eqref{eq:flambdadbwigner} into~\eqref{eq:fclambdadbi} and is given by
\begin{equation}
	\begin{split}
        f_{C}(c) = &\frac{5\cdot\lnL{2}2^{c}}{\pi\sigma_{\mdg}\lnL{10}\left(2^{c}-1\right)}\\
                   &\sqrt{4-\frac{1}{\sigma^{2}_{\mdg}}\left(\frac{10}{\lnL{10}}\lnL{\frac{2^{c}-1}{\SNR}}-\mu_{\lambda_{\dB}}\right)^{2}},
    \end{split}
    \label{eq:fcwigner}
\end{equation}
and the \gls*{cdf}, developed in further detail in \appref{app:e}, is given by
\begin{equation}
	\begin{split}
		F_{C}(c) = &\frac{1}{2}+\left(\frac{5}{\pi\sigma_{\mdg}\lnL{10}}\lnL{\frac{2^c-1}{\SNR}}-\frac{\mu_{\lambda_{\dB}}}{2\pi\sigma_{\mdg}}\right)\\
                   &\sqrt{1-\left(\frac{5}{\sigma_{\mdg}\lnL{10}}\lnL{\frac{2^c-1}{\SNR}}-\frac{\mu_{\lambda_{\dB}}}{2\sigma_{\mdg}}\right)^{2}}\\
                   &+\frac{1}{\pi}\arcsinL{\frac{5}{\sigma_{\mdg}\lnL{10}}\lnL{\frac{2^c-1}{\SNR}}-\frac{\mu_{\lambda_{\dB}}}{2\sigma_{\mdg}}},
	\end{split}
	\label{eq:Fc}
\end{equation}
where both distributions are limited in the following support
\begin{equation}
    \begin{split}
	    \logL{2}{\SNR\cdot 10^{\frac{\mu_{\lambda_{\dB}}-2\sigma_{\mdg}}{10}}+1} \leq c \\
	    \leq \logL{2}{\SNR\cdot 10^{\frac{\mu_{\lambda_{\dB}}+2\sigma_{\mdg}}{10}}+1}.
    \end{split}
\end{equation}

\begin{figure}[!t]
    \centering
    \includegraphics{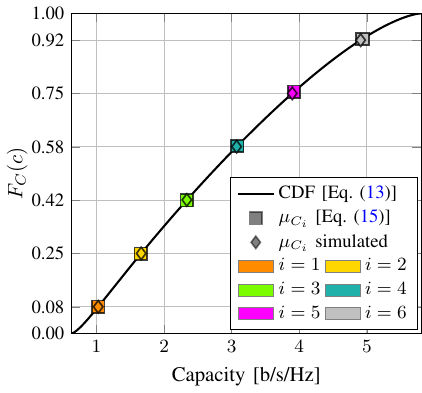}
    \caption{Capacity \gls*{cdf} and ${\mu_{C_{i}}}$ estimation for ${D=6}$, ${\SNR=10~\dB}$, and ${\sigma_{\mdg}=5~\dB}$.}
    \label{fig:CDF}
\end{figure}

Assuming that the individual \gls*{pdf}s are symmetric and have minimal overlap, the values of $\mu_{C_{i}}$ can be obtained from the \gls*{cdf} of the ensemble as
\begin{equation}
	\mu_{C_{i}}=F^{-1}_{C}\left(\frac{i-1/2}{D}\right),
    \label{eq:ucicdf}
\end{equation}
where $F^{-1}_{C}$ is the inverse function of the capacity \gls*{cdf}.

An assessment of the \gls*{cdf}-derived approximation can be found in Fig.~\ref{fig:CDF}, for a system with $D=6$, $\SNR = 10~\dB$ and $\sigma_{\mdg} = 5~\dB$, showing excellent agreement.

The values of $\sigma_{C_{i}}$ can then be obtained from the ensemble \gls*{pdf} and $\mu_{C_{i}}$, such as in \eqref{eq:sigmalambdadb}, yielding
\begin{equation}
	\sigma_{C_{i}} = \frac{1}{D\sqrt{2\pi}f_{C}(\mu_{C_{i}})}.
	\label{eq:sigmaciwigner}
\end{equation}

Differences between the predictions generated by the Wigner semicircular distribution and the \gls*{gue} spectral are expected. The Wigner semicircular distribution flattens the local maxima of the \gls*{gue} spectral distribution. However, the divergences fade as the mode count increases. In Fig.~\ref{fig:Marginals}, the predictions for $D=12$, $D=20$ and $D=40$ are obtained based on the Wigner semicircle distribution. A significant $\sigma_{C_{i}}$ error is observed at $D=12$, attributed to the divergence of the Wigner semicircular distribution from the actual distribution. This error diminishes as the number of modes increases.

% Section: 3
\section{Total Capacity Distribution $C_T$}
\label{sec:correlation}

Owing to correlations between $\lambda_{i}$ variables, the individual distributions are not sufficient to model the statistical parameters of the total capacity. As seen in Fig.~\ref{fig:MargHist}, the distributions of per-mode individual capacities closely approximate normal distributions. Therefore, in this work, we also propose to approximate the total capacity PDF by a multivariate normal distribution. A sum of individual elements of a multivariate normal distribution has a normal distribution given by
\begin{equation}
	f_{C_{T}}(c_{T})=\frac{1}{\sigma_{C_{T}}\sqrt{2\pi}}e^{-\frac{1}{2}\left(\frac{c_{T}-\mu_{C_{T}}}{\sigma_{C_{T}}}\right)^{2}},
\end{equation}
where the variance $\sigma_{C_{T}}^{2}$ and mean $\mu_{C_{T}}$ can be obtained, respectively, by
\begin{align}
    \begin{split}
       \sigma^{2}_{C_{T}}&=\sum_{i,j=1,1}^{D,D}\sigma_{C_{i}}\sigma_{C_{j}}\rho_{i,j}\\
        \mu_{C_{T}}&=\sum_{i=1}^{D}\mu_{C_{i}}.
    \end{split}
    \label{eq:fctvalues}
\end{align}

The average $\mu_{C_{T}}$ presented in (18) holds true given our assumption that the per-mode capacity is Gaussian distributed. Alternatively, an exact solution for $\mu_{C_{T}}$, derived directly from the modal gains ensemble distribution, has already been proposed in~\cite[Eq.~(10)]{ho2011mode},~\cite[Eq.~(12)]{mello2020impact}
\begin{equation}
    \mu_{C_{T}}=D\int_{-\infty}^{\infty}\logL{2}{1+\SNR\cdot10^{\frac{\sigma_{\mdg}x+\mu_{\lambda_{\dB}}}{10}}}\widetilde{f}_{\lambda_{\dB}}(x)dx,
    \label{eq:fcintmean}
\end{equation}
where $\widetilde{f}_{\lambda_{\dB}}(x)$ is the unitary variance, zero mean $\lambda_{\dB}$ distribution.

In (\ref{eq:fctvalues}), $\rho_{i,j}$ is the correlation coefficient between the $i^{th}$ and $j^{th}$ capacity \gls*{pdf}s. The correlation coefficient is related to the covariance by
\begin{equation}
	\rho_{i,j} = \frac{\cov{C_{i},C_{j}}}{\sigma_{C_{i}}\sigma_{C_{j}}}.
\end{equation}

\begin{figure}[t]
    \centering
    \includegraphics{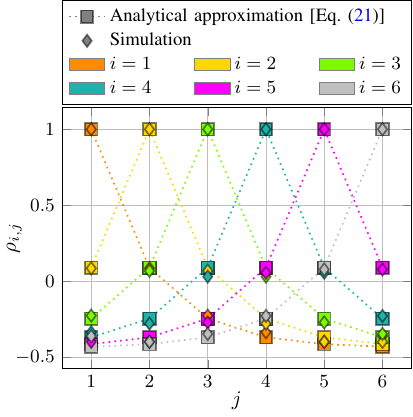}
    \caption{Correlation for ${D=6}$, ${\SNR=10~\dB}$, and ${\sigma_{\mdg}=5~\dB}$. Dotted lines connect the analytical correlation values for visualization purposes.}
    \label{fig:Correlation}
\end{figure}

\begin{table*}[b]
    \centering
    \caption{Correlation function coefficients. The results were obtained with  $10^{-10}$  $\gamma_{0}$ error tolerance and $10^{4}$  $\gamma_{1}$ iterations.}
    \label{tab:Coefficients}
    \TabularCoefficients
\end{table*}

\begin{figure*}[t]
    \centering
    \includegraphics{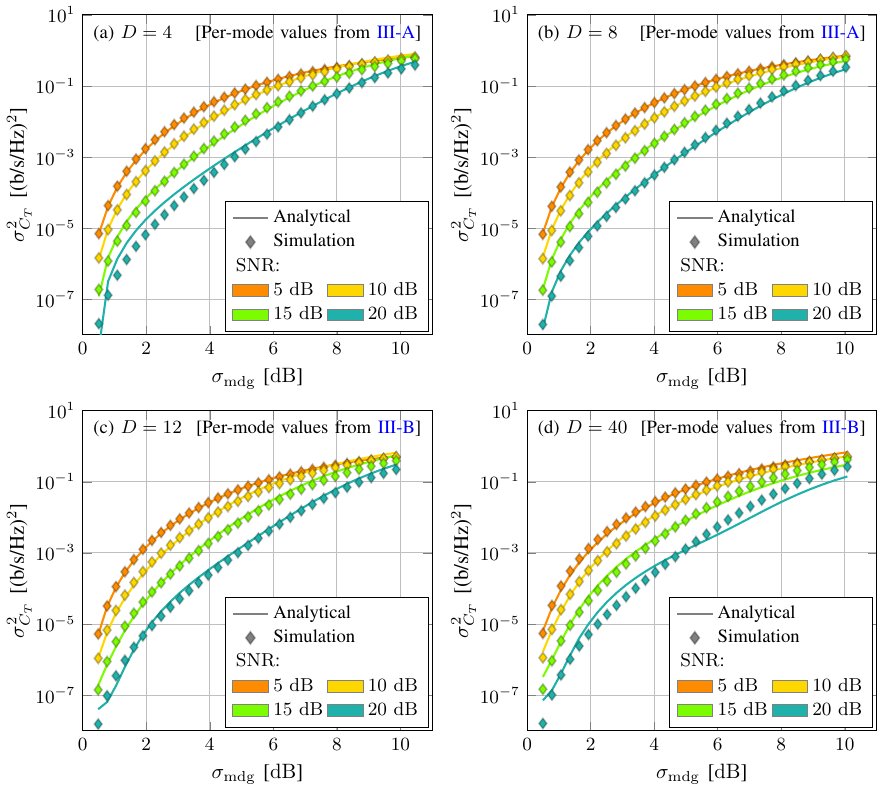}
    \caption{Estimation of the total capacity variance with the analytical model compared to simulated values. For $4$ (a) and $8$ modes (b), the analytical method relies on the individual statistical values derived from the \gls*{gue} spectral distribution, as discussed in Section~\ref{ssec:gue_spectral_distribution}. For $12$ (c) and $40$ (d) modes, values are derived from the Wigner semicircular distribution, as discussed in Section~\ref{ssec:wigner_semicircular_distribution}.}
    \label{fig:TotalVar}
\end{figure*}

\begin{figure}[!t]
    \centering
    \includegraphics{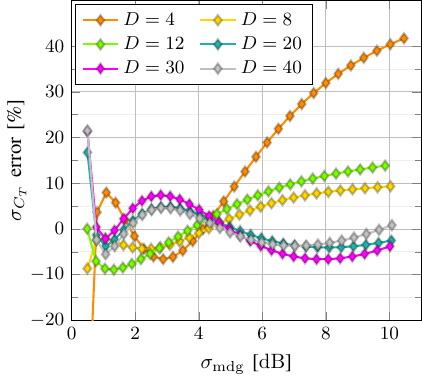}
    \caption{Approximation error for ${\SNR=10~\dB}$. For comparison fairness, all the per-mode individual values are obtained from the Wigner semicircular \gls*{cdf} method.}
    \label{fig:LimitConvergence}
\end{figure}

An exact equation for $\rho_{i,j}$ is unknown. However, as observed in Fig.~\ref{fig:Correlation}, a suitable empirical approximation is given by
\begin{equation}
	\rho_{i,j} = e^{-|i-j|}+\left(e^{-|i-j|}-1\right)\left(\gamma_{0}+\gamma_{1}\cdot\sigma_{\mdg}^{2.75}\right),
    \label{eq:correlation}
\end{equation}
where the coefficients $\gamma_{0}$ and $\gamma_{1}$ are specific to a given mode count and \gls*{snr} value. In this paper, we numerically obtain the values of $\gamma_{0}$ and $\gamma_{1}$ such that the \gls*{msle} between analytical and simulation capacity variance is minimized, i.e.,
\begin{equation}
    \left(\gamma_{0},\gamma_{1}\right)=\mathrm{argmin}_{\gamma_{0},\gamma_{1}}\left(\mathbb{E}\left\{\left[\lnL{\widehat{\sigma}^{2}_{C_{T}}}-\lnL{\sigma^{2}_{C_{T}}}\right]^{2}\right\}\right),
\end{equation}
where $\widehat{\sigma}^{2}_{C_{T}}$ is the total capacity variance obtained via simulation.

The methodology to obtain the coefficients is performed iteratively until a tolerance threshold is reached. The algorithm implemented in this work firstly updates $\gamma_{0}$ until the error for the lowest considered $\sigma_{\mdg}$ is smaller than a defined tolerance. The value of $\gamma_{1}$ is updated considering the \gls*{msle} for the entire evaluated $\sigma_{\mdg}$ range, with small corrections to $\gamma_{0}$ if the resulting approximation function is not monotonic increasing. We select the logarithmic error because the capacity variance scales exponentially with $\sigma_{\mdg}$ when measured in decibels. The coefficients for a practical range of mode counts and \gls*{snr} are given in Table~\ref{tab:Coefficients}.

\begin{figure}[!t]
    \centering
    \includegraphics{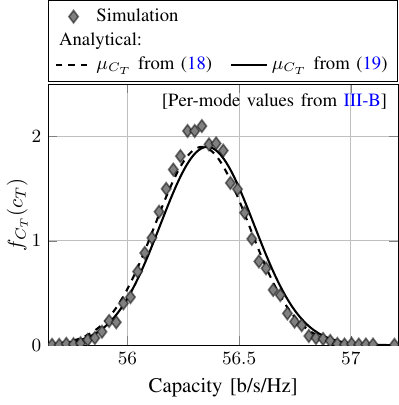}
    \caption{Total Gaussian \gls*{pdf} for ${D=20}$, ${\SNR=10~\dB}$, and ${\sigma_{\mdg}=5~\dB}$.}
    \label{fig:TotalHist}
\end{figure}

An evaluation of the analytical method is presented in Fig.~\ref{fig:TotalVar}. The results indicate an accurate variance approximation for \gls*{snr} values lower than $20~\dB$ for all evaluated mode counts. For $\textrm{SNR}=20~\dB$, there is a higher divergence, particularly with the Wigner semicircular \gls*{cdf} method.
The source of estimation discrepancy likely stems from the per-mode individual distributions estimated from the capacity \gls{cdf}, as well as approximations made in the correlation function. Fig.~\ref{fig:LimitConvergence} evaluates the accuracy of the 
method based on Wigner semicircular \gls*{cdf} for different values of $\sigma_{\mdg}$ and mode counts. The results evidence accurate estimates for mode counts higher than 20. Fig.~\ref{fig:TotalHist} compares the simulated total capacity distribution with the analytical distributions produced with the Wigner semicircular \gls*{cdf} method. The presented Gaussian approximation exhibits a good agreement with the simulated curve.

% Section: Diversity
\section{Impact of frequency diversity}
\label{sec:freqdiv}

\begin{figure}[!h]
    \centering
    \includegraphics{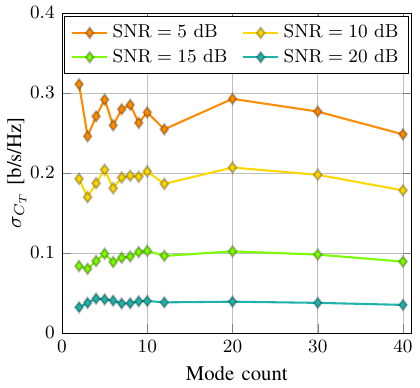}
    \caption{Total capacity deviation per mode count for ${\sigma_{\mdg}=5~\dB}$ from the analytical equations.}
    \label{fig:VarPerMode}
\end{figure}

In \gls*{mdg}-impaired systems, the random coupling among modes turns the capacity into a random variable, eventually generating outages. As demonstrated in \cite{ho2011frequency}, modal dispersion generates a frequency-dependent channel frequency response that mitigates the impact of deep fades. This effect, known as frequency diversity, benefits the outage capacity~\cite{arik2014diversity}. As seen in Fig.~\ref{fig:VarPerMode}, the total capacity standard deviation has a weak dependence on the mode count, in agreement with~\cite{arik2014diversity}. This effect has been called mode diversity in \cite{mello2020impact}.
According to the effect of mode diversity, the outage probability is more significant in lower mode counts, as fewer paths are available to mitigate outages.

While this paper has so far only discussed a narrowband channel, as presented in~\cite{ho2011frequency}, modal dispersion induces frequency-dependent mode gains. In this case, the total capacity in a given bandwidth can be calculated by dividing the spectrum into $N$ narrowband frequency bins with uncorrelated modal gains, i.e. $N$ subchannels, and adding the corresponding capacities~\cite{mello2020impact}. In this model, the channel response of a given frequency bin is independent of that in adjacent bins~\cite{ho2011frequency}. As the coherence bandwidth of the modal gains is in the order of the inverse standard deviation of the group delay $\sigma_{\text{GD}}$~\cite{ho2011frequency}, $N$ is linearly proportional to $\sigma_{\text{GD}}$. Indeed, $N$ can be approximated by the given bandwidth multiplied by $\sigma_{\text{GD}}$ \cite{ho2011frequency,arik2014diversity} Considering the distributions presented in~Section~\ref{sec:marginaldistribution}, the averaged capacity variance of the $i^{th}$ mode, considering $N$ independent frequency bins is given by
\begin{equation}
    \overline{\sigma}^{2}_{C_{i}}=\frac{\sigma^{2}_{C_{i}}}{N}.
    \label{eq:freqdiversity}
\end{equation}

Substituting the individual capacity variances in~\eqref{eq:fctvalues} with~\eqref{eq:freqdiversity}, we see that the average total capacity variance is reduced by the same rate of $1/N$.

% Section: Case study
\section{Case Study}
\label{sec:casestudy}

Assuming strong-mode-coupling, the formulas presented in this paper can be used for system design if the accumulated \gls*{mdg} and optical \gls*{snr} are known to the operator. Estimation techniques for both parameters are discussed in~\cite{ospina2022mdg} and~\cite{ospina2023digital}. As a case study, we calculate the outage capacity of a 3-mode polarization multiplexed link ($D=6$) with $\sigma_{\mdg}=5~\dB$ and $\SNR=10~\dB$, for an outage probability  $p_{\mathrm{out}}=1\%$. We use the \gls*{gue} spectral distribution method presented in Section~\ref{ssec:gue_spectral_distribution}.

The mean logarithmic power gain, considering an amplification scheme with total power control, is obtained from~\eqref{eq:mulambdadb} by numeric integration as

\begin{equation}
    \mu_{\lambda_{\dB}}=-2.609~\dB.
\end{equation}
%-2.6089806

The $\lambda_{\dB}$ distribution coefficients can be obtained from~\cite[Table 1]{ho2011mode} or derived analytically, as detailed in~\appref{app:a}. They are given by 
\begin{equation}
    \begin{split}
        \alpha_{\lambda_{\dB},6}=&\sqrt{\frac{14}{\pi}}\\
        \boldsymbol{\beta}_{\lambda_{\dB},6}=&\left[\frac{322}{3125},~\frac{4557}{1250},~\frac{-17493}{625},~\right. \\
        &\left. ~\frac{256221}{3125},~\frac{-259308}{3125},~\frac{453789}{15625}\right].
    \end{split}
\end{equation}

The individual capacities mean and standard deviation are calculated from~\eqref{eq:ulambdaguederivative} and \eqref{eq:sigmalambdadb}, respectively, and are given by
\begin{equation}
    \begin{split}
        \boldsymbol{\mu}_{C}\hspace{-1mm}=&[1.022,~1.653,~2.330,~3.067,~3.887,~4.865],~[b/s/Hz]\\
        \boldsymbol{\sigma}_{C}\hspace{-1mm}=&[0.192,~0.202,~0.217,~0.238,~0.276,~0.362],~[b/s/Hz],
    \end{split}
\end{equation}
where the values of $\boldsymbol{\mu}_{C}$ are obtained numerically, from the odd-index crossings of~\eqref{eq:ulambdaguederivative} along the $x$ axis.

The individual capacity distribution for the given parameters is presented in Fig.~\ref{fig:MargHist}.

The correlation coefficients are obtained from~\eqref{eq:correlation}, considering $\gamma_{0}=0.43513127$ and $\gamma_{1}=3.758373\cdot 10^{-5}$, as provided in table~\ref{tab:Coefficients}, and are given in matrix form as
\newcommand{\minus}{\hspace{-2.5mm}-}
\begin{equation}
    \boldsymbol{\rho}\hspace{-0.5mm}=\hspace{-1.5mm}
    \begin{bmatrix}
        \hspace{2.5mm}1 & 0.091 & \minus 0.244 & \minus 0.367 & \minus 0.412 & \minus 0.429 \\
        \hspace{2.5mm}0.091 & 1 & 0.091 & \minus 0.244 & \minus 0.367 & \minus 0.412 \\
        -0.244 & 0.091 & 1 & 0.091 & \minus 0.244 & \minus 0.367 \\
        -0.367 & \minus 0.244 & 0.091 & 1 & 0.091 & \minus 0.244 \\
        -0.412 & \minus 0.367 & \minus 0.244 & 0.091 & 1 & 0.091 \\
        -0.429 & \minus 0.412 & \minus 0.367 & \minus 0.244 & 0.091 & 1 \\
    \end{bmatrix}
\end{equation}

The total capacity statistical parameters are obtained from~\eqref{eq:fctvalues} and are given by
\begin{equation}
    \begin{split}
        \mu_{C_{T}}= \hspace{1mm}& 16.825,~[b/s/Hz]\\
        \sigma_{C_{T}}= \hspace{1mm}& 0.181,~[b/s/Hz].
    \end{split}
    \label{eq:extotalcap}
\end{equation}
%16.8246342
%0.1808305
%\mu_{C_{T},int} = 16.9516309

\begin{figure}[t]
    \centering
    \includegraphics{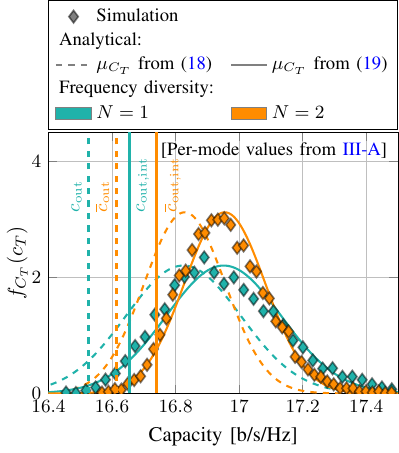}
    \caption{Total Gaussian \gls*{pdf} for ${D=6}$, ${\SNR=10~\dB}$, and ${\sigma_{\mdg}=5~\dB}$. For the frequency diverse case, we consider 2 uncorrelated frequency bins, where we show the capacity average over the subchannels. The estimated outage capacities consider an outage probability of ${p_{\mathrm{out}}=1\%}$.}
    \label{fig:OutageHist}
\end{figure}

The total capacity distribution for the given parameters can be seen in Fig.~\ref{fig:OutageHist}.

Considering the Gaussian model, the outage capacity $c_{\mathrm{out}}$ can be obtained from the Gaussian \gls*{cdf} and the outage probability as
\begin{equation}
    c_{\mathrm{out}}=F_{C_{T}}^{-1}(p_{\mathrm{out}})=\sqrt{2}\sigma_{C_{T}}\mathrm{erf}^{-1}\left(2p_{\mathrm{out}}-1\right)+\mu_{C_{T}},
\end{equation}
where $\mathrm{erf}^{-1}(\cdot)$ is the inverse error function.

For the defined parameters the outage capacity is
\begin{equation}
    c_{\mathrm{out}}=16.527,~[b/s/Hz].
\end{equation}
%16.5271723

Considering the total capacity mean presented in~\eqref{eq:fctvalues}, the outage capacity is marginally underestimated. Using~\eqref{eq:fcintmean}, the outage capacity is more accurately estimated as
\begin{equation}
    c_{\mathrm{out},\mathrm{int}}=16.654,~[b/s/Hz].
\end{equation}
%16.6541690

In the presence of frequency diversity of $N=2$, the average total capacity deviation is reduced and is given by
\begin{equation}
    \overline{\sigma}_{C_{T}}=0.128,~[b/s/Hz].
\end{equation}
%=0.1278665

Considering the mean capacity given by~\eqref{eq:fctvalues}, the reduced variance increases the average outage capacity to
\begin{equation}
    \overline{c}_{\mathrm{out}}=16.614,~[b/s/Hz].
\end{equation}
%16.6142968
%16.7412935306

We validate the frequency diverse case against simulated values in Fig.~\ref{fig:OutageHist}. For each Monte-Carlo trial, we simulate 2 uncorrelated channel matrices $\mathbf{H}$, and average the capacity between both subchannels.

% Section: Conclusion
\section{Conclusion}
\label{sec:conclusion}

We propose statistical models that approximate the individual per-mode and total capacity variances for strongly-coupled \gls*{sdm} transmission systems. The model is based on the Gaussian assumption for the individual per-mode capacity. The total capacity distribution is modeled as a joint multivariate normal distribution given by a linear combination of the individuals.
The individual per-mode capacity distributions are derived using two methods: one based on the \gls*{gue} spectral distribution for channel gains, and the other based on the Wigner semicircular distribution. The correlation between per-mode individual capacities is addressed by an empirical analytical model with fitting coefficients obtained by numerical simulations. The analytical results exhibit a suitable agreement with simulations in the investigated scenarios. The narrowband results are extended to arbitrary bandwidths resorting to existing theory on frequency diversity.

% Appendices
\appendices
\def\sectionautorefname{appendix}

% Eigenvalues distribution Appendix
\section{Distribution of the ensemble of $\lambda_{\dB}$ eigenvalues}
\label{app:a}

As described in~\cite{ho2011statistics}, the logarithm power gain $\lambda_{\dB}$ follows the fixed-trace \gls*{gue}. The joint \gls*{pdf} of the \gls*{gue} spectral distribution is given by~\cite[Eq.~(3.3.7)]{mehta2004random}
\begin{equation}
    f_{\boldsymbol{\lambda}}(\boldsymbol{\lambda})=\alpha_{\boldsymbol{\lambda},D}\exp\left(-\sum_{k=1}^{D}\lambda_{k}^{2}\right)\prod_{\substack{j,i\in[1,D]\\i<j}}(\lambda_{i}-\lambda_{j})^{2},
\end{equation}
with $\lambda_{1}<\lambda_{2}<\cdots<\lambda_{D}$, and where $\alpha_{\boldsymbol{\lambda},D}$ is a normalization factor, such that the \gls*{pdf} integral is unitary.

The fixed-trace \gls*{gue} has the additional constraint that the sum of the eigenvalues is constant. As pointed out by~\cite{rosenzweig1963graphical}, the fixed-trace is a property of stable systems. Considering a \gls*{gue} matrix $\mathbf{A}\in\mathbb{C}^{D\times D}$, the fixed-trace condition specifies that
\begin{equation}
    \text{tr}(\mathbf{A})=\sum_{i=1}^{D}\lambda_{i}=\kappa,
\end{equation}
where $\text{tr}(\cdot)$ represents the matrix trace and $\kappa$ is a constant.

The ensemble spectral distribution, related to any eigenvalue $\lambda$ in  set $\lambda_{i}, i\in[1, D]$, is obtained from the $k$-point correlation function in the case of $k=1$, defined in~\cite[III]{dyson1962statistical} and \cite[Eq.~(6.1.2)]{mehta2004random}.

The fixed-trace solution of the ensemble spectral distribution normalized to zero mean is developed in detail in~\cite[From (S.3) to (S.15)]{ho2011statistics}. Defining an auxiliary variable $t$ where its $n^{th}$ order polynomial is expressed in terms of $\lambda$ as
\begin{equation}
    t^{n}\xmapsto{}\left[(-1)^{n}H_{n}\left(\lambda\sqrt{\frac{D(D+1)}{2}}\right)\right],
\end{equation}
the fixed-trace solution normalized to unitary variance is given by
\begin{equation}
    f_{\lambda}(\lambda)=\frac{\sqrt{D+1}}{D\sqrt{2\pi}}e^{-\frac{(D+1)}{2}\lambda^{2}}\sum^{D-1}_{k=0}\frac{1}{2^{k}k!}H^{2}_{k}\left(\frac{t}{2\sqrt{D-1}}\right),
    \label{eq:tlambda}
\end{equation}
where $H_{n}(\cdot)$ is the physicist's Hermite polynomial given by
\begin{equation}
    H_{n}(x)=(-1)^{n}e^{x^{2}}\frac{d^{n}}{dx^{n}}e^{-x^{2}}=n!\sum_{k=0}^{\lfloor n/2\rceil}(-1)^{k}\frac{(2x)^{n-2k}}{k!(n-2k)!},
    \label{eq:flamcomplex}
\end{equation}
where $\lfloor\cdot\rceil$ denotes rounding to the nearest integer.

Expanding the Hermite polynomial in~\eqref{eq:flamcomplex} and replacing the auxiliary variable $t$ as defined in~\eqref{eq:tlambda}, the logarithmic gain distribution, with proper mean and standard deviation scaling, can be rewritten as~\eqref{eq:flambdadb}.

As pointed out by~\cite[I]{wigner1955characteristic}, at the limit $D\rightarrow\infty$ the distribution given in~\eqref{eq:flamcomplex} converges to
\begin{equation}
    \lim_{D\rightarrow\infty}f_{\lambda}(\lambda)=\frac{1}{2\pi}\sqrt{4-\lambda^{2}},~2\leq\lambda\leq 2,
\end{equation}
and with proper mean and standard deviation scaling, is rewritten as~\eqref{eq:flambdadbwigner}.

% Eigenvalues mean Appendix
\section{Equation for the mean $\mu_{\lambda_{\dB}}$ of the logarithmic scale eigenvalues $\lambda_{\dB}$}
\label{app:b}

Given $\lambda_{\dB}=10\cdot\logL{10}{\lambda}$, the linear scale eigenvalue distribution is related to the $\lambda_{\dB}$ distribution by
\begin{equation}
    f_{\lambda}(\lambda)=\frac{10}{\lnL{10}\lambda}f_{\lambda_{\dB}}\left(10\cdot\logL{10}{\lambda}\right),
\end{equation}
and the $\lambda$ mean is given by
\begin{equation}
    \mu_{\lambda}=\int_{-\infty}^\infty\frac{10}{\lnL{10}}\widehat{f}_{\lambda_{\dB}}\left(10\cdot\logL{10}{\lambda}-\mu_{\lambda_{\dB}}\right)d\lambda,
\end{equation}
where $\widehat{f}_{\lambda_{\dB}}(\cdot)$ is the $\lambda_{\dB}$ distribution shifted by $\mu_{\lambda_{\dB}}$, resulting in a zero mean \gls*{pdf}. Using $x=10\cdot\logL{10}{\lambda}-\mu_{\lambda_{\dB}}$ and $dx=\lnL{10}10^{\frac{x+\mu_{\lambda_{\dB}}}{10}}/10~d\lambda$, after some manipulation we have
\begin{equation}
    \mu_{\lambda_{\dB}}=-10\cdot\mathrm{log}_{10}\left[\int_{-\infty}^{\infty}\frac{10^{\frac{x}{10}}}{\mu_{\lambda}}\widehat{f}_{\lambda_{\dB}}\left(x\right)dx\right].
\end{equation}

% GUE derivative Appendix
\section{Approximation of $\mu_{\lambda_{\dB,i}}$ from the GUE derivative}
\label{app:c}

We approximate the means of individual eigenvalues $\lambda_{\dB,i}$ by local maxima of $f_{\lambda_{\dB}}(\lambda_{\dB})$. Given that the \gls*{gue} distribution is positive definite and is zero at the limit $\lambda_{\dB}\rightarrow -\infty$, defining properties of probabilistic distributions, the first $\lambda_{\dB,i}$ value for which $f'_{\lambda_{\dB}}(\lambda_{\dB})=0$ is a local maximum, and the subsequent root is a local minimum. Maxima and minima alternate in a way such that odd solutions return local maxima. This can be demonstrated by guaranteeing that $f''_{\lambda_{\dB}}(\lambda_{\dB})<0$ for the $\lambda_{\dB}$ values that fulfill $f'_{\lambda_{\dB}}(\lambda_{\dB})=0$. Expanding on the first derivative from (\ref{eq:flambdadb}), utilizing $\widehat{\lambda}_{\dB}=\lambda_{\dB}-\mu_{\lambda_{\dB}}$, yields 
\begin{equation}
	\begin{aligned}
		&f'_{\widehat{\lambda}_{\dB}}(\widehat{\lambda}_{\dB}) = 0 \\
        &\frac{d}{d\widehat{\lambda}_{\dB}}\left[\frac{\alpha_{\lambda_{\dB},D}}{\sigma_{\mdg}}e^{-\frac{(D+1)}{2}\frac{\widehat{\lambda}_{\dB}^{2}}{\sigma_{\mdg}^{2}}}\sum_{k=0}^{D-1}\beta_{\lambda_{\dB},D,k}\frac{\widehat{\lambda}_{\dB}^{2k}}{\sigma_{\mdg}^{2k}}\right] = 0\\
		&\frac{d}{d\widehat{\lambda}_{\dB}}\left[e^{-\frac{(D+1)}{2}\frac{\widehat{\lambda}_{\dB}^{2}}{\sigma_{\mdg}^{2}}}\sum_{k=0}^{D-1}\beta_{\lambda_{\dB},D,k}\frac{\widehat{\lambda}_{\dB}^{2k}}{\sigma_{\mdg}^{2k}}\right] = 0.
	\end{aligned}
\end{equation}
Setting $g(\widehat{\lambda}_{\dB})=\sum_{k=0}^{D-1}\beta_{\lambda_{\dB},D,k}\frac{\widehat{\lambda}_{\dB}^{2k}}{\sigma_{\mdg}^{2k}}$ and $\delta_{0}=\frac{D+1}{2\sigma_{\mdg}^{2}}$,
\begin{equation}
    \begin{aligned}
        &\frac{d}{d\widehat{\lambda}_{\dB}}\left[e^{-\delta_{0}\lambda_{\dB}^{2}}g(\widehat{\lambda}_{\dB})\right] = 0\\
        &g'(\widehat{\lambda}_{\dB})e^{-\delta_{0}\widehat{\lambda}_{\dB}^{2}} - 2\delta_{0}\widehat{\lambda}_{\dB}e^{-\delta_{0}\widehat{\lambda}_{\dB}^{2}}g(\widehat{\lambda}_{\dB}) = 0\\
        &g'(\widehat{\lambda}_{\dB}) - 2\delta_{0}\widehat{\lambda}_{\dB}g(\widehat{\lambda}_{\dB}) = 0,
    \end{aligned}
    \label{eq:b2}
\end{equation}
where $g'(\widehat{\lambda}_{\dB})$ is given by
\begin{equation}
    g'(\widehat{\lambda}_{\dB}) = \sum_{k=0}^{D-1}\frac{2k\beta_{\lambda_{\dB},D,k}}{\sigma_{\mdg}^{2}}\widehat{\lambda}_{\dB}^{2k-1}.
    \label{eq:b3}
\end{equation}
Applying~\eqref{eq:b3} into~\eqref{eq:b2}, and replacing  $\widehat{\lambda}_{\dB}$ by $\mu_{\lambda_{\dB,i}}-\mu_{\lambda_{\dB}}$ yields~\eqref{eq:ulambdaguederivative} after some manipulation.

% Capacity CDF Appendice
\section{Approximation of $\mu_{C_{i}}$  from the derivative of the individual capacities distribution}
\label{app:d}

The distribution for individual capacities  is provided in \eqref{eq:fclambdadbi}.   We now assume that a Gaussian distribution can approximate the distribution of individual capacities. If this approximation holds, the mean capacity corresponds to the point where the derivative of the probability density function (PDF) is zero, i.e., the mean can be obtained from the solution to $f'_{C_{i}}(c_{i})=0$.  To derive this result we consider the following auxiliary elements
\begin{equation}
    \begin{split}
        \delta_{1} &= \frac{10\cdot\lnL{2}}{\sigma_{\lambda_{\dB,i}}\lnL{10}\sqrt{2\pi}},\\
        h_{0}(c_{i}) &= \frac{1}{\sigma_{\lambda_{\dB,i}}}\left[\frac{10}{\lnL{10}}\lnL{\frac{2^{c_{i}}-1}{\SNR}}-\mu_{\lambda_{\dB,i}}\right],\\
        h_{1}(c_{i}) &= e^{\frac{-h_{0}^{2}(c_{i})}{2}},\\
        h_{2}(c_{i}) &= 2^{c_{i}}h_{1}(c_{i}).
    \end{split}
    \label{eq:c1}
\end{equation}

Applying~\eqref{eq:flambdadbi} into~\eqref{eq:fclambdadbi}, and applying the substitutions provided in~\eqref{eq:c1}, yields
\begin{equation}
    f_{C_{i}}(c_{i}) = \frac{\delta_{1}}{\left(2^{c_{i}}-1\right)}h_{2}(c_{i}),
\end{equation}
with the derivative given by
\begin{equation}
    f'_{C_{i}}(c_{i}) = \frac{\delta_{1}}{\left(2^{c_{i}}-1\right)^{2}}\left[h'_{2}(c_{i})\left(2^{c_{i}}-1\right)-h_{2}(c_{i})\lnL{2}2^{c_{i}}\right].
    \label{eq:c3}
\end{equation}

The derivatives of the auxiliary functions in~\eqref{eq:c1} are given by
\begin{equation}
    \begin{split}
        h'_{0}(c_{i}) &= \frac{\delta_{1}\sqrt{2\pi}\cdot 2^{c_{i}}}{\left(2^{c_{i}}-1\right)}\\
        h'_{1}(c_{i}) &= -h_{0}(c_{i})h'_{0}(c_{i})e^{\frac{-h_{0}^{2}(c_{i})}{2}}\\
        h'_{2}(c_{i}) &= 2^{c_{i}}\left(h'_{1}(c_{i})+\lnL{2}h_{1}(c_{i})\right).
    \end{split}
\end{equation}
Replacing these terms into~\eqref{eq:c3} results in
\begin{equation}
    \begin{split}
        f'_{C_{i}}(c_{i}) &= \frac{\delta_{1}2^{c_{i}}}{\left(2^{c_{i}}-1\right)^{2}}[\left(h'_{1}(c_{i})+\lnL{2}h_{1}(c_{i})\right)\left(2^{c_{i}}-1\right)\\
                          &~~ -2^{c_{i}}\lnL{2}h_{1}(c_{i})]\\
                          &= \frac{\delta_{1}2^{c_{i}}}{\left(2^{c_{i}}-1\right)^{2}}\left[h'_{1}(c_{i})\left(2^{c_{i}}-1\right)-\lnL{2}h_{1}(c_{i})\right]\\
                          &= \frac{-\delta_{1}2^{c_{i}}e^{\frac{-h_{0}^{2}(c_{i})}{2}}\lnL{2}}{\left(2^{c_{i}}-1\right)^{2}}\left[\frac{\delta_{1}h_{0}(c_{i})2^{c_{i}}\sqrt{2\pi}}{\lnL{2}}+1\right].
    \end{split}
    \label{eq:c5}
\end{equation}

For $f'_{C_{i}}(c_{i})=0$, ~\eqref{eq:c5} can be simplified. Given that the capacity is positive, the first multiplicative terms of~\eqref{eq:c5} are always non-zero, as expressed by
\begin{equation}
    \frac{-\delta_{1}2^{c_{i}}e^{\frac{-h_{0}^{2}(c_{i})}{2}}\lnL{2}}{\left(2^{c_{i}}-1\right)^{2}} < 0,\ \forall c_{i}\in(0,\infty),
\end{equation}
and therefore, can be disregarded.

After simplifications, replacing the auxiliary terms from \eqref{eq:c1} into~\eqref{eq:c5}, and replacing $c_{i}$ with $\mu_{C_{i}}$, results in~\eqref{eq:mucigue}.

% Capacity CDF Appendix
\section{Capacity CDF given the semicircular $\lambda_{\dB}$ distribution}
\label{app:e}

The capacity \gls*{cdf} considering the ensemble of eigenvalues can be obtained by integrating~\eqref{eq:fcwigner}. Considering $\delta_{2}=\sigma_{\mdg}\lnL{10}/10$, the \gls*{cdf} is given by
\begin{equation}
    \begin{aligned}
	F_{C}(c)=& \int_{c_{0}}^{c}f_{C}(x)dx \\
            =& \frac{\lnL{2}}{2\pi\delta_{2}}\int_{c_{0}}^{c}\frac{2^{x}}{2^{x}-1}\\
            &\sqrt{4-\left(\frac{1}{\delta_{2}}\lnL{\frac{2^{x}-1}{\SNR}}-\frac{\mu_{\lambda_{\dB}}}{\sigma_{\mdg}}\right)^{2}}dx,
    \end{aligned}
\end{equation}
where $c_{0}=\mathrm{log}_{2}(\SNR\cdot 10^{\frac{1}{10}\left(\mu_{\lambda_{\dB}}-2\sigma_{\mdg}\right)}+1)$ is the smallest value of $c$ such that $f_{C}(c)$ is strictly real, obtainable from the lowest real-valued boundary of the terms inside the square root. Applying the following variable substitutions
\begin{equation}
    \begin{aligned}
        y &= \arcsinL{\frac{1}{2\delta_{2}}\lnL{\frac{2^{x}-1}{\SNR}}-\frac{\mu_{\lambda_{\dB}}}{2\sigma_{\mdg}}}\\
        y_{0} &= \arcsinL{\frac{1}{2\delta_{2}}\lnL{\frac{2^{c_{0}}-1}{\SNR}}-\frac{\mu_{\lambda_{\dB}}}{2\sigma_{\mdg}}}=\frac{-\pi}{2}\\
        y_{f} &= \arcsinL{\frac{1}{2\delta_{2}}\lnL{\frac{2^{c}-1}{\SNR}}-\frac{\mu_{\lambda_{\dB}}}{2\sigma_{\mdg}}}\\
        dy &= \frac{\lnL{2}\cdot 2^{x}}{\delta_{2}\left(2^{x}-1\right)\sqrt{4-4\cdot\sinPowL{2}{y}}}dx,
    \end{aligned}
    \label{eq:d2}
\end{equation}
results in
\begin{equation}
    \begin{aligned}
	F_{C}(c) = \frac{2}{\pi}\int_{y_{0}}^{y_{f}}\cosPowL{2}{y}dy = \frac{1}{\pi}\left[\cosL{y}\sinL{y}+y\right]_{y=y_{0}}^{y_{f}}.
    \end{aligned}
    \label{eq:d3}
\end{equation}

Applying the necessary substitutions from \eqref{eq:d2} into \eqref{eq:d3} results in the \gls*{cdf} given in~\eqref{eq:Fc}.

% Acknowledgment
%\section*{Acknowledgment}

% No acknowledgment

% Bibliography
\bibliographystyle{IEEEtran}
\bibliography{references}

\begin{thebibliography}{10}
\providecommand{\url}[1]{#1}
\csname url@samestyle\endcsname
\providecommand{\newblock}{\relax}
\providecommand{\bibinfo}[2]{#2}
\providecommand{\BIBentrySTDinterwordspacing}{\spaceskip=0pt\relax}
\providecommand{\BIBentryALTinterwordstretchfactor}{4}
\providecommand{\BIBentryALTinterwordspacing}{\spaceskip=\fontdimen2\font plus
\BIBentryALTinterwordstretchfactor\fontdimen3\font minus
  \fontdimen4\font\relax}
\providecommand{\BIBforeignlanguage}[2]{{%
\expandafter\ifx\csname l@#1\endcsname\relax
\typeout{** WARNING: IEEEtran.bst: No hyphenation pattern has been}%
\typeout{** loaded for the language `#1'. Using the pattern for}%
\typeout{** the default language instead.}%
\else
\language=\csname l@#1\endcsname
\fi
#2}}
\providecommand{\BIBdecl}{\relax}
\BIBdecl

\bibitem{morioka2009new}
T.~Morioka, ``{New generation optical infrastructure technologies:“EXAT
  initiative” towards 2020 and beyond},'' in \emph{2009 14th OptoElectronics
  and Communications Conference}.\hskip 1em plus 0.5em minus 0.4em\relax IEEE,
  2009, pp. 1--2.

\bibitem{essiambre2012capacity}
R.-J. Essiambre and R.~W. Tkach, ``{Capacity trends and limits of optical
  communication networks},'' \emph{Proceedings of the IEEE}, vol. 100, no.~5,
  pp. 1035--1055, 2012.

\bibitem{winzer2012optical}
P.~J. Winzer, ``{Optical networking beyond WDM},'' \emph{IEEE Photonics
  Journal}, vol.~4, no.~2, pp. 647--651, 2012.

\bibitem{richardson2013space}
D.~J. Richardson, J.~M. Fini, and L.~E. Nelson, ``{Space-division multiplexing
  in optical fibres},'' \emph{Nature photonics}, vol.~7, no.~5, pp. 354--362,
  2013.

\bibitem{saitoh2016multicore}
K.~Saitoh and S.~Matsuo, ``{Multicore fiber technology},'' \emph{Journal of
  lightwave technology}, vol.~34, no.~1, pp. 55--66, 2016.

\bibitem{puttnam2021space}
B.~J. Puttnam, G.~Rademacher, and R.~S. Lu{\'\i}s, ``{Space-division
  multiplexing for optical fiber communications},'' \emph{Optica}, vol.~8,
  no.~9, pp. 1186--1203, 2021.

\bibitem{cristiani2022roadmap}
I.~Cristiani, C.~Lacava, G.~Rademacher, B.~J. Puttnam, R.~S. Lu{\`\i}s,
  C.~Antonelli, A.~Mecozzi, M.~Shtaif, D.~Cozzolino, D.~Bacco \emph{et~al.},
  ``{Roadmap on multimode photonics},'' \emph{Journal of Optics}, vol.~24,
  no.~8, p. 083001, 2022.

\bibitem{rademacher202010}
G.~Rademacher, B.~J. Puttnam, R.~S. Lu{\'\i}s, J.~Sakaguchi, W.~Klaus, T.~A.
  Eriksson, Y.~Awaji, T.~Hayashi, T.~Nagashima, T.~Nakanishi \emph{et~al.},
  ``{10.66 peta-bit/s transmission over a 38-core-three-mode fiber},'' in
  \emph{Optical Fiber Communication Conference}.\hskip 1em plus 0.5em minus
  0.4em\relax Optica Publishing Group, 2020, pp. Th3H--1.

\bibitem{rademacher2021peta}
G.~Rademacher, B.~J. Puttnam, R.~S. Lu{\'\i}s, T.~A. Eriksson, N.~K. Fontaine,
  M.~Mazur, H.~Chen, R.~Ryf, D.~T. Neilson, P.~Sillard \emph{et~al.},
  ``{Peta-bit-per-second optical communications system using a standard
  cladding diameter 15-mode fiber},'' \emph{Nature Communications}, vol.~12,
  no.~1, p. 4238, 2021.

\bibitem{rademacher20221}
G.~Rademacher, R.~S. Lu{\'\i}s, B.~J. Puttnam, N.~K. Fontaine, M.~Mazur,
  H.~Chen, R.~Ryf, D.~T. Neilson, D.~Dahl, J.~Carpenter \emph{et~al.}, ``{1.53
  peta-bit/s c-band transmission in a 55-mode fiber},'' in \emph{2022 European
  Conference on Optical Communication (ECOC)}.\hskip 1em plus 0.5em minus
  0.4em\relax IEEE, 2022, pp. 1--4.

\bibitem{winzer2014mode}
P.~J. Winzer, H.~Chen, R.~Ryf, K.~Guan, and S.~Randel, ``{Mode-dependent loss,
  gain, and noise in MIMO-SDM systems},'' in \emph{2014 The European Conference
  on Optical Communication (ECOC)}.\hskip 1em plus 0.5em minus 0.4em\relax
  IEEE, 2014, pp. 1--3.

\bibitem{winzer2011mimo}
P.~J. Winzer and G.~J. Foschini, ``{MIMO capacities and outage probabilities in
  spatially multiplexed optical transport systems},'' \emph{Optics express},
  vol.~19, no.~17, pp. 16\,680--16\,696, 2011.

\bibitem{yamamoto1993observation}
S.~Yamamoto, N.~Edagawa, H.~Taga, Y.~Yoshida, and H.~Wakabayashi,
  ``{Observation of BER degradation due to fading in long-distance optical
  amplifier system},'' \emph{Electronics Letters}, vol.~29, no.~2, pp.
  209--210, 1993.

\bibitem{taylor1993observation}
M.~Taylor, ``{Observation of new polarization dependence effect in long haul
  optically amplified system},'' \emph{IEEE photonics technology letters},
  vol.~5, no.~10, pp. 1244--1246, 1993.

\bibitem{pilipetskii2006performance}
A.~N. Pilipetskii and E.~A. Golovchenko, ``{Performance fluctuations in
  submarine WDM systems},'' \emph{Journal of lightwave technology}, vol.~24,
  no.~11, pp. 4208--4214, 2006.

\bibitem{shtaif2004polarization}
M.~Shtaif and A.~Mecozzi, ``{Polarization-dependent loss and its effect on the
  signal-to-noise ratio in fiber-optic systems},'' \emph{IEEE Photonics
  Technology Letters}, vol.~16, no.~2, pp. 671--673, 2004.

\bibitem{nelson2011statistics}
L.~E. Nelson, C.~Antonelli, A.~Mecozzi, M.~Birk, P.~Magill, A.~Schex, and
  L.~Rapp, ``{Statistics of polarization dependent loss in an installed
  long-haul WDM system},'' \emph{Optics express}, vol.~19, no.~7, pp.
  6790--6796, 2011.

\bibitem{velazquez2018scaling}
A.~M. Vel{\'a}zquez-Ben{\'\i}tez, J.~E. Antonio-L{\'o}pez, J.~C.
  Alvarado-Zacar{\'\i}as, N.~K. Fontaine, R.~Ryf, H.~Chen,
  J.~Hern{\'a}ndez-Cordero, P.~Sillard, C.~Okonkwo, S.~G. Leon-Saval
  \emph{et~al.}, ``{Scaling photonic lanterns for space-division
  multiplexing},'' \emph{Scientific reports}, vol.~8, no.~1, p. 8897, 2018.

\bibitem{wada2018recent}
M.~Wada, T.~Sakamoto, S.~Aozasa, T.~Yamamoto, and K.~Nakajima, ``{Recent
  progress on SDM amplifiers},'' in \emph{2018 European Conference on Optical
  Communication (ECOC)}.\hskip 1em plus 0.5em minus 0.4em\relax IEEE, 2018, pp.
  1--3.

\bibitem{wada2018cladding}
M.~Wada, T.~Sakamoto, T.~Yamamoto, S.~Aozasa, S.~Nozoe, Y.~Sagae, K.~Tsujikawa,
  and K.~Nakajima, ``{Cladding pumped randomly coupled 12-core erbium-doped
  fiber amplifier with low mode-dependent gain},'' \emph{journal of lightwave
  technology}, vol.~36, no.~5, pp. 1220--1225, 2018.

\bibitem{antonelli2015modeling}
C.~Antonelli, A.~Mecozzi, M.~Shtaif, and P.~J. Winzer, ``{Modeling and
  performance metrics of MIMO-SDM systems with different amplification schemes
  in the presence of mode-dependent loss},'' \emph{Optics Express}, vol.~23,
  no.~3, pp. 2203--2219, 2015.

\bibitem{antonelli2019stokes}
C.~Antonelli, A.~Mecozzi, M.~Shtaif, N.~K. Fontaine, H.~Chen, and R.~Ryf,
  ``{Stokes-space analysis of modal dispersion of SDM fibers with
  mode-dependent loss: Theory and experiments},'' \emph{Journal of Lightwave
  Technology}, vol.~38, no.~7, pp. 1668--1677, 2019.

\bibitem{ho2011mode}
K.-P. Ho and J.~M. Kahn, ``{Mode-dependent loss and gain: statistics and effect
  on mode-division multiplexing},'' \emph{Optics Express}, vol.~19, no.~17, pp.
  16\,612--16\,635, Aug 2011.

\bibitem{ho2011frequency}
------, ``{Frequency diversity in mode-division multiplexing systems},''
  \emph{Journal of Lightwave Technology}, vol.~29, no.~24, pp. 3719--3726,
  2011.

\bibitem{mello2020impact}
D.~A. Mello, H.~Srinivas, K.~Choutagunta, and J.~M. Kahn, ``{Impact of
  polarization-and mode-dependent gain on the capacity of ultra-long-haul
  systems},'' \emph{Journal of Lightwave Technology}, vol.~38, no.~2, pp.
  303--318, 2020.

\bibitem{melo2024on}
H.~V. Melo, R.~S. Ospina, and D.~A. Mello, ``{On the Gaussian Assumption for
  the Capacity of MDG-Impaired Systems},'' in \emph{2024 IEEE Photonics
  Conference (IPC). Accepted for publication}.\hskip 1em plus 0.5em minus
  0.4em\relax IEEE, 2024.

\bibitem{paulraj2003introduction}
A.~Paulraj, R.~Nabar, and D.~Gore, \emph{{Introduction to space-time wireless
  communications}}.\hskip 1em plus 0.5em minus 0.4em\relax Cambridge university
  press, 2003.

\bibitem{juarez2012perspectives}
A.~A. Juarez, C.~A. Bunge, S.~Warm, and K.~Petermann, ``{Perspectives of
  principal mode transmission in mode-division-multiplex operation},''
  \emph{Optics express}, vol.~20, no.~13, pp. 13\,810--13\,824, 2012.

\bibitem{ho2013mode}
K.-P. Ho, J.~M. Kahn, I.~Kaminow, T.~Li, and A.~Willner, ``{Mode coupling and
  its impact on spatially multiplexed systems},'' \emph{Optical Fiber
  Telecommunications VI}, vol.~17, pp. 1386--1392, 2013.

\bibitem{efimov2014spatial}
A.~Efimov, ``{Spatial coherence at the output of multimode optical fibers},''
  \emph{Optics express}, vol.~22, no.~13, pp. 15\,577--15\,588, 2014.

\bibitem{ho2011statistics}
K.-P. Ho and J.~M. Kahn, ``{Statistics of Group Delays in Multimode Fiber With
  Strong Mode Coupling},'' \emph{Journal of Lightwave Technology}, vol.~29,
  no.~21, pp. 3119--3128, Aug 2011.

\bibitem{wigner1955characteristic}
E.~P. Wigner, ``{Characteristic vectors of bordered matrices with infinite
  dimensions},'' \emph{Annals of Mathematics}, vol.~62, no.~3, pp. 548--564,
  1955.

\bibitem{arik2014diversity}
S.~{\"O}. Ar{\i}k and J.~M. Kahn, ``{Diversity-multiplexing tradeoff in
  mode-division multiplexing},'' \emph{Optics Letters}, vol.~39, no.~11, pp.
  3258--3261, 2014.

\bibitem{ospina2022mdg}
R.~S. Ospina, M.~v.~d. Hout, S.~van~der Heide, J.~van Weerdenburg, R.~Ryf,
  N.~K. Fontaine, H.~Chen, R.~Amezcua-Correa, C.~Okonkwo, and D.~A. Mello,
  ``{MDG and SNR estimation in SDM transmission based on artificial neural
  networks},'' \emph{Journal of Lightwave Technology}, vol.~40, no.~15, pp.
  5021--5030, 2022.

\bibitem{ospina2023digital}
R.~S. Ospina, D.~A. Mello, L.~Zischler, R.~S. Lu{\'\i}s, B.~J. Puttnam,
  H.~Furukawa, M.~van~den Hout, S.~van~der Heide, C.~Okonkwo, R.~Ryf
  \emph{et~al.}, ``{Digital Signal Processing for MDG Estimation in Long-Haul
  SDM Transmission},'' \emph{Journal of Lightwave Technology}, 2023.

\bibitem{mehta2004random}
M.~L. Mehta, \emph{{Random matrices}}.\hskip 1em plus 0.5em minus 0.4em\relax
  Elsevier, 2004.

\bibitem{rosenzweig1963graphical}
M.~L. Rosenzweig and R.~H. MacArthur, ``{Graphical representation and stability
  conditions of predator-prey interactions},'' \emph{The American Naturalist},
  vol.~97, no. 895, pp. 209--223, 1963.

\bibitem{dyson1962statistical}
F.~J. Dyson, ``{Statistical theory of the energy levels of complex systems},''
  \emph{Journal of Mathematical Physics}, vol.~3, no.~1, pp. 140--156, 1962.

\end{thebibliography}

\end{document}